\documentclass[aps,prl,preprint,superscriptaddress]{revtex4}

\usepackage{graphicx,latexsym}

\begin{document}


\title{Solid-liquid phase coexistence and structural transitions in palladium clusters}


\author{D. Schebarchov}
\affiliation{School of Chemical and Physical Sciences,
Victoria University of Wellington, New Zealand}
\author{S. C. Hendy}
\affiliation{School of Chemical and Physical Sciences,
Victoria University of Wellington, New Zealand}
\affiliation{MacDiarmid Institute for Advanced Materials
and Nanotechnology, Industrial Research Ltd, Lower Hutt, New Zealand}



\date{\today}

\begin{abstract}
We use molecular dynamics with an embedded atom potential to study the behavior of palladium nanoclusters near the melting point in the microcanonical ensemble. We see transitions from both fcc and decahedral ground state structures to icosahedral structures prior to melting over a range of cluster sizes. In all cases this transition occurs during solid-liquid phase coexistence and the mechanism for the transition appears to be fluctuations in the molten fraction of the cluster and subsequent recrystallization into the icosahedral structure. 
\end{abstract}


\maketitle



One of the goals of nanoparticle science is to determine the stable structure of a particle at a given size and temperature \cite{Baletto05}. Metals that are fcc in their bulk phase are known to exhibit a variety of thermodynamically stable non-crystalline cluster structures including icosahedra \cite{Ino67} and decahedra \cite{Marks84}. However, on experimental timescales it is frequently the kinetics, rather than the thermodynamics, that determine the structure of a nanoparticle and this can obscure the true thermodynamically stable state \cite{Baletto05}. On the other hand, the kinetic stability of a range of cluster structures does mean that we may potentially be able to control cluster structure, and through this, their properties. For instance, it was recently demonstrated that one could induce an icosahedral to decahedral change in structure which remained stable upon cooling, by annealing gold particles near their melting point \cite{Koga04}. Thus the study of solid-solid structural transitions in metal nanoclusters is important both for developing a deeper understanding cluster thermodynamics and kinetics, and for the technological spin-offs that may follow if we can tailor nanocluster structure.   

In general, structural transitions at high temperatures in solid clusters will be driven by entropic effects \cite{Baletto05}. For example, changes in structure driven by favorable vibrational entropy from fcc to decahedra or icosahedra have been predicted theoretically \cite{Doye01}, and seen in simulations \cite{Cleveland98}. However, such transitions can also be driven by energetics. Simulations of a 1415-atom nickel nanoparticle revealed a transition from an icosahedral structure (the minimum energy {\em solid} structure at this size) to decahedral structure whilst in a partially melted state \cite{Hendy05b}. This transformation appeared to be driven by an energetic preference of the melt to wet the (100)-facets of the decahedron rather than favorable entropy.     

One approach for identifying such solid-solid transitions is to map out cluster caloric curves. A solid-solid transition may be distinguished by a sharp change in internal energy of the nanocluster at the transition temperature. There are several experimental approaches to measuring cluster caloric curves, such as photofragmentation \cite{Haberland01} and multicollision dissociation \cite{Breaux03}. Both techniques measure the internal energy of mass-selected clusters at a given temperature by fragmentation. Using the latter technique Breaux et al \cite{Breaux05} have seen premelting features in the caloric curves of small aluminium clusters. However, interpreting these features is difficult: are they solid-solid transitions or surface melting, or a combination of both? 

The study of structural transitions and surface melting using molecular dynamics (MD) simulations is potentially a useful tool for resolving such ambiguities. Such computational experiments can help us either directly interpret the premelting features, or at least catalogue the variety of behavior likely to be seen in such experiments. Furthermore comparison with the calorimetry experiments provides an invaluable test of the simulation methodology \cite{Noya05}.   
   
Here we report on MD simulations of premelting transitions in Palladium nanoclusters. Pd clusters show potential for a variety of technological applications, including catalysis \cite{Beck02}, hydrogen storage \cite{Zuttel03} and for use as hydrogen detection devices \cite{Walter02}. Indeed, it is thought that in the presence of hydrogen, fcc Pd clusters can undergo transitions to icosahedral structures \cite{Pundt02,Pundt04}. Here we report solid-solid transitions, in the absence of hydrogen, from both fcc and decahedral structures to icosahedral structures near the melting point in Pd nanoclusters over a range of sizes. Further we find that dynamic and static solid-liquid coexistence\cite{Berry93} plays an important role in the kinetics of these solid-solid transitions.

There have been a number of molecular dynamics studies of nanometer-sized Pd clusters \cite{Barreteau00,Baletto02,Pundt02,Lopez04}. Here we will use an embedded atom method (EAM) potential for Pd \cite{Foiles86}, which was used in the study by J. L. Rodr\'{i}guez-L\'{o}pez et al \cite{Lopez04}. Where they overlap, these studies generally agree that there is a cross-over in energetic preference from icosahedral to decahedral structures at 561 atoms, and a cross-over between decahedral and fcc structures at sizes of several thousand atoms. The study by Pundt et al \cite{Pundt02} found that a 2057-atom cuboctahedral structure underwent a transition to an icosahedral structure at elevated temperatures, and that this new icosahedral structure was energetically favored over the fcc cuboctahedron. However, cuboctahedra are rarely the optimal structural form for fcc metal clusters, so while this is not necessarily inconsistent with other studies, it is not strong evidence of a preference for icosahedral structures over fcc structures at high temperatures. 

\begin{figure}
\resizebox{\columnwidth}{!}{\includegraphics{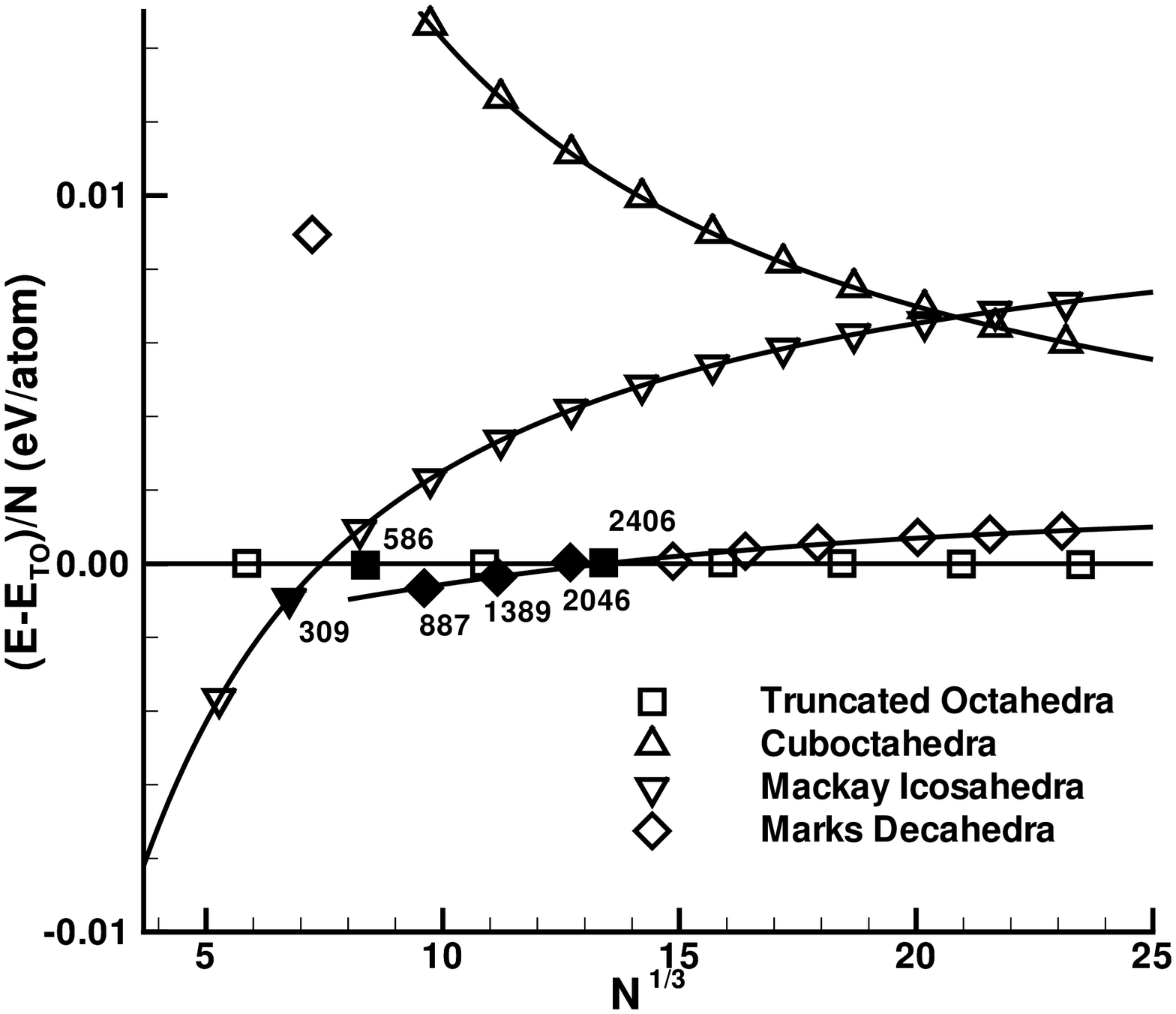}}
\caption{\label{fig1} Comparison of the relaxed zero temperature energies of Mackay icosahedra, Marks decahedra, cuboctahedra and truncated octahedra for palladium clusters. Energies are given relative to a fit (cubic in $N^{1/3}$) 
to the energies of the truncated octahedra sequence. Filled symbols indicate clusters for which caloric curves were generated.}
\end{figure}

Figure~\ref{fig1} shows the energetics of several structural sequences relative to the energy of the closed-shell truncated octahedron (TO) sequence using the EAM potential \cite{Foiles86}. We see that icosahedra are energetically favored up to 309-atoms. The 586-atom TO is stable, but then the decahedral sequence becomes stable from 887-atoms up to 2046-atoms. From this point the TO sequence is favored. Note that the cuboctahedra are less stable than the icosahedra until the cluster size reaches 8217 atoms. Our calculations here are consistent with previous work, although we predict a lower than usual threshold for the crossover between decahedral and fcc structures.    

We have constructed microcanonical caloric curves for a selection of stable structures (distinguished by filled symbols in Figure~\ref{fig1}) using the following procedure: at each fixed total energy the cluster was equilibrated for 150000 time steps (where $\Delta t$ = 2.7 fs) and then the kinetic energy was averaged over a further 150000 steps to obtain a temperature. Uniform scaling of the kinetic energy, with an energy increment of 0.5 meV/atom, was used to adjust the total energy between simulations. To identify and characterize solid-liquid coexistence, we followed Cleveland et al \cite{Cleveland94}, using the bimodality of the distribution of diffusion coefficients to distinguish solid and liquid atoms. Structural changes were identified using a CNA-based \cite{CNA93} classification scheme for clusters \cite{Hendy01, Hendy02}.

In the microcanonical ensemble, we generally expect to see phase coexistence in clusters prior to melting, although thermodynamic arguments suggest that there is a lower bound on cluster sizes beyond which phase coexistence becomes unstable \cite{Hendy05a}. Indeed, a distinct threshold for static phase coexistence was seen in molecular dynamics simulations of Pb icosahedra \cite{Hendy05a}, although no such threshold was seen in similar studies of Ag, Cu or Ni clusters \cite{Hendy05c}. Here we clearly observed phase coexistence in all the clusters we examined, and saw no evidence for a Pb-like size threshold for coexistence.  

\begin{figure}
\resizebox{\columnwidth}{!}{\includegraphics{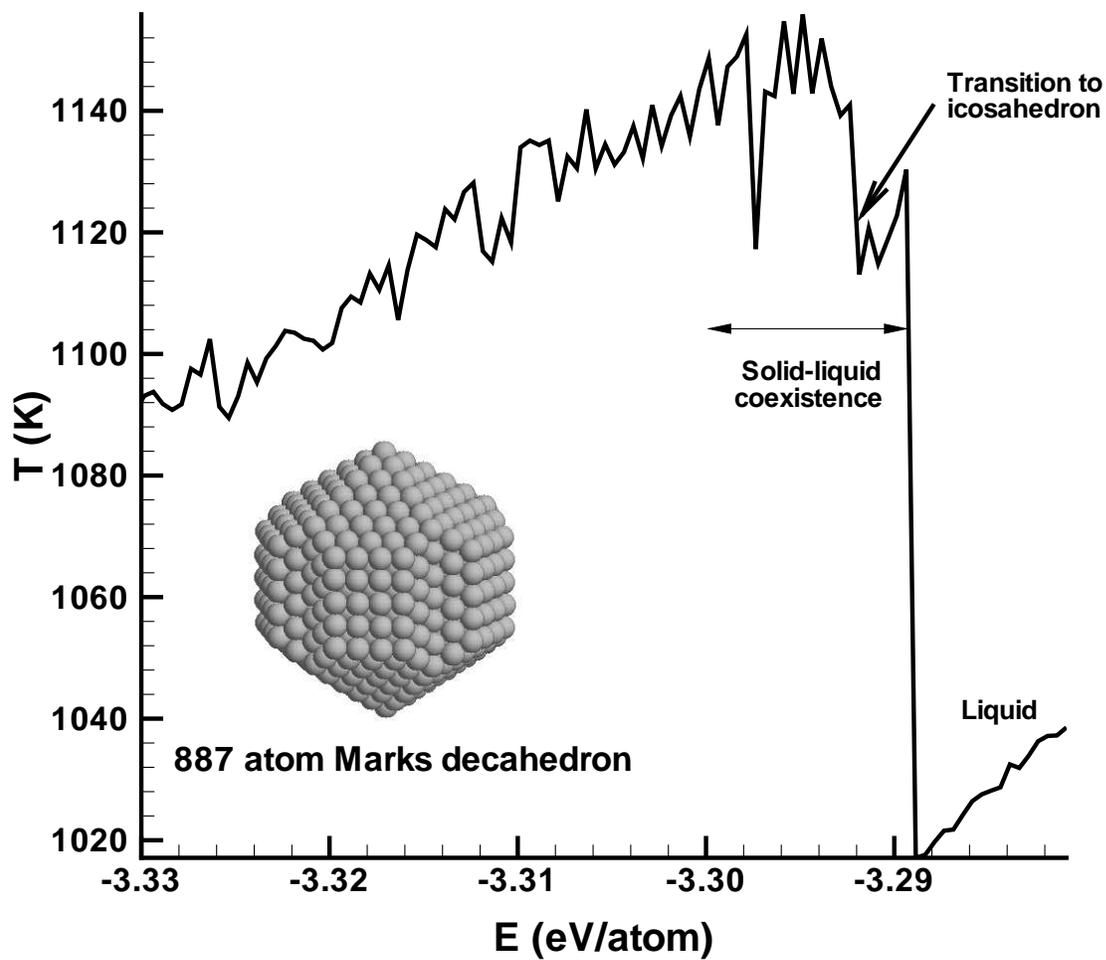}}
\caption{\label{fig2}The caloric curve for the 887-atom Marks decahedron near the melting point.}
\end{figure}

\begin{figure}
\resizebox{\columnwidth}{!}{\includegraphics{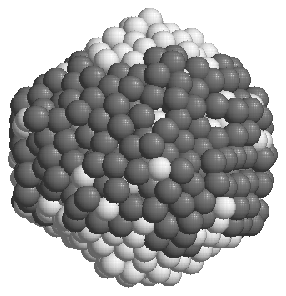} \includegraphics{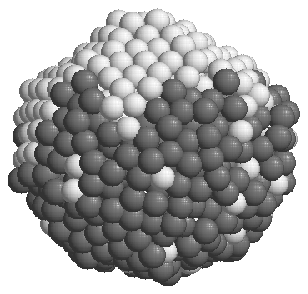} \includegraphics{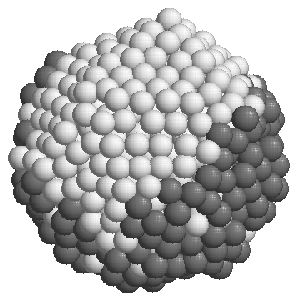}}
\caption{\label{fig2a} The structure of the coexisting decahedral structure (left), the almost molten intermediate structure (center) and the coexisting icosahedral structure (right). Atoms with a darker shade have been identified as liquid.}
\end{figure}

Figure~\ref{fig2} shows the caloric curve for the 887-atom Marks decahedron. We observe the onset of solid-liquid coexistence at a total energy of $E=-3.300$ eV/atom and full melting at $E=-3.288$ eV/atom. The melting point of a cluster is generally reduced, compared to that of the bulk, due to the favorable surface energy of the liquid \cite{Buffat76}. Here we see the onset of coexistence at $T=1150$ K, well below the bulk melting temperature of $T_c=1830$ K. In the microcanonical ensemble, full melting is expected to occur when the coexisting state becomes unstable at some size-dependent critical liquid fraction \cite{Nielsen94}. Prior to melting, the largest liquid fraction we observe is approximately 0.6. Note that during coexistence the (100) facets and reentrant edges of the decahedron preferentially melt, leaving the (111) facets exposed as shown in the leftmost snapshot in figure~\ref{fig2a}.

However, between the onset of coexistence and full melting, we see another transition at $E=-3.292$ eV/atom from the partially melted decahedron to a new structure which is best characterized as a partially melted icosahedron. This structural transition is indicated in the caloric curve by a drop in temperature of cluster as kinetic energy is traded for potential energy.  

\begin{figure}
\resizebox{\columnwidth}{!}{\includegraphics{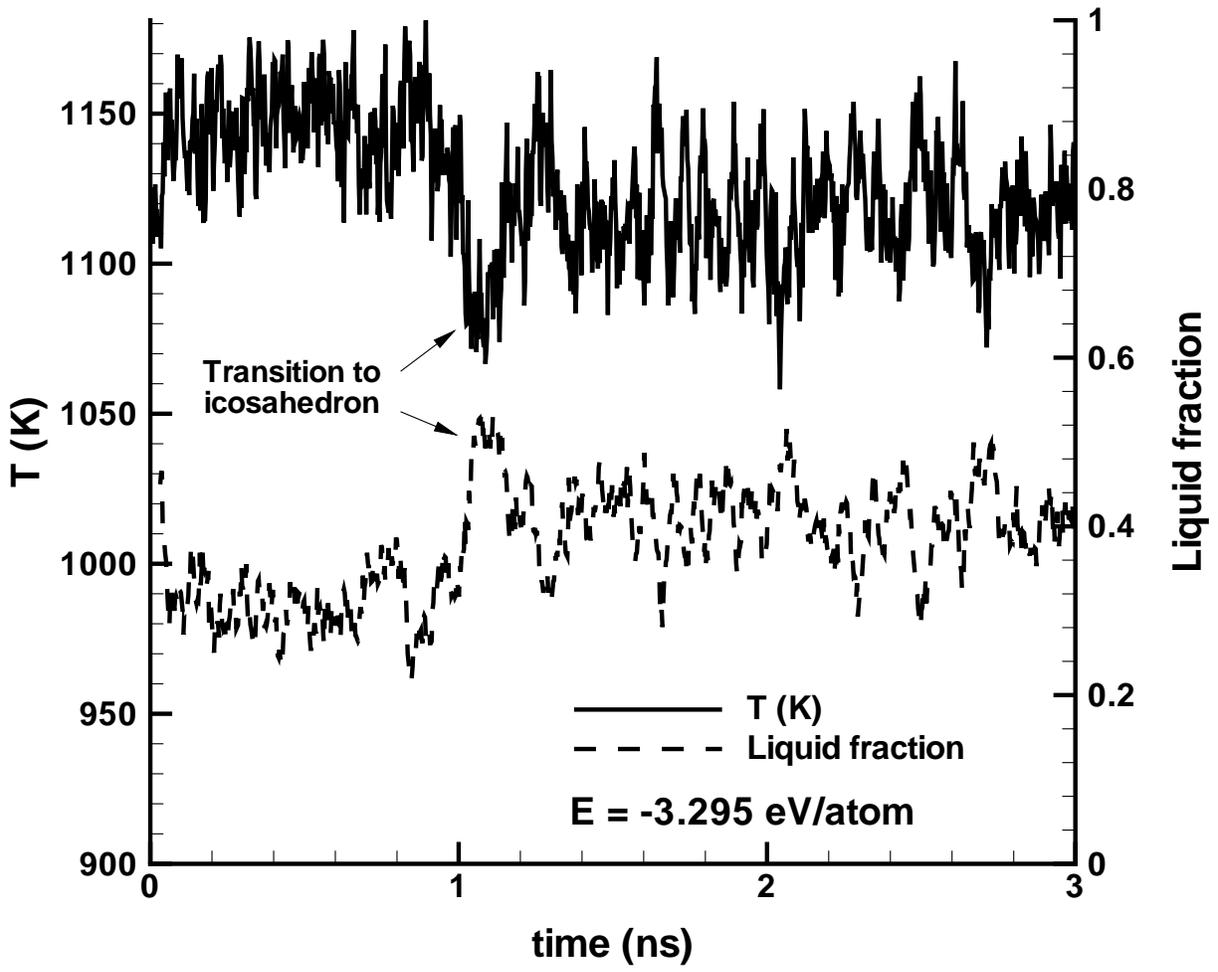}}
\caption{\label{fig3}The temperature and liquid fraction of an 887-atom cluster at $E=-3.295$ eV/atom initially prepared in a decahedral structure.}
\end{figure}

To examine this transition further, we performed several longer simulations of the (initially) decahedral 887-atom cluster at energies that lie in the coexistence region. Figure~\ref{fig3} shows the evolution of the temperature and the liquid fraction for the cluster at $E=-3.295$ eV/atom (an energy just above that where we observe the transition in the caloric curve). Here we observe that a transition occurs at approximately 1 ns into the simulation, where the temperature drops by 70 K and the liquid fraction jumps from 0.3 to 0.5. Almost immediately the temperature recovers by 40 K and the liquid fraction stabilises at a level of 0.4. Snapshots from before, during and after this transition, are shown in figure~\ref{fig2a}.  

The snapshots reveal that at the transition, the coexisting decahedron largely melts, leaving only one fivefold grouping of fcc tetrahedra intact at one of the fivefold apices of the decahedron. Subsequently, recrystallization into a new structure occurs coinciding with previously noted drop in liquid fraction and rise in temperature. This new structure is commensurate with the surviving fivefold apex of the decahedron, but contains new fivefold apices sharing (111)-facets and facet edges. Quenching this structure, either rapidly or relatively slowly, results in recrystallization of the cluster into a full icosahedral structure. Thus the new structure is best described as a partial icosahedron coexisting with the melt.

Note that the presence of the melt does not energetically favor the icosahedral structure (judging by the increase in cluster potential energy, or equivalently, the drop in temperature), although it clearly plays a role in the kinetics of the transition. This makes an estimate of the transition energy difficult as the presence of the melt presumably lowers the energy barrier for the transition, making the timescale for the transition accessible by our simulations only during coexistence. However, a 15 ns simulation of the coexisting decahedron at $E=-3.300$ eV/atom (where the liquid fraction was approximately 0.25) did not find a similar transition, providing some reassurance that the transition energy indeed falls in the range of energies where coexistence occurs. 

\begin{figure}
\resizebox{\columnwidth}{!}{\includegraphics{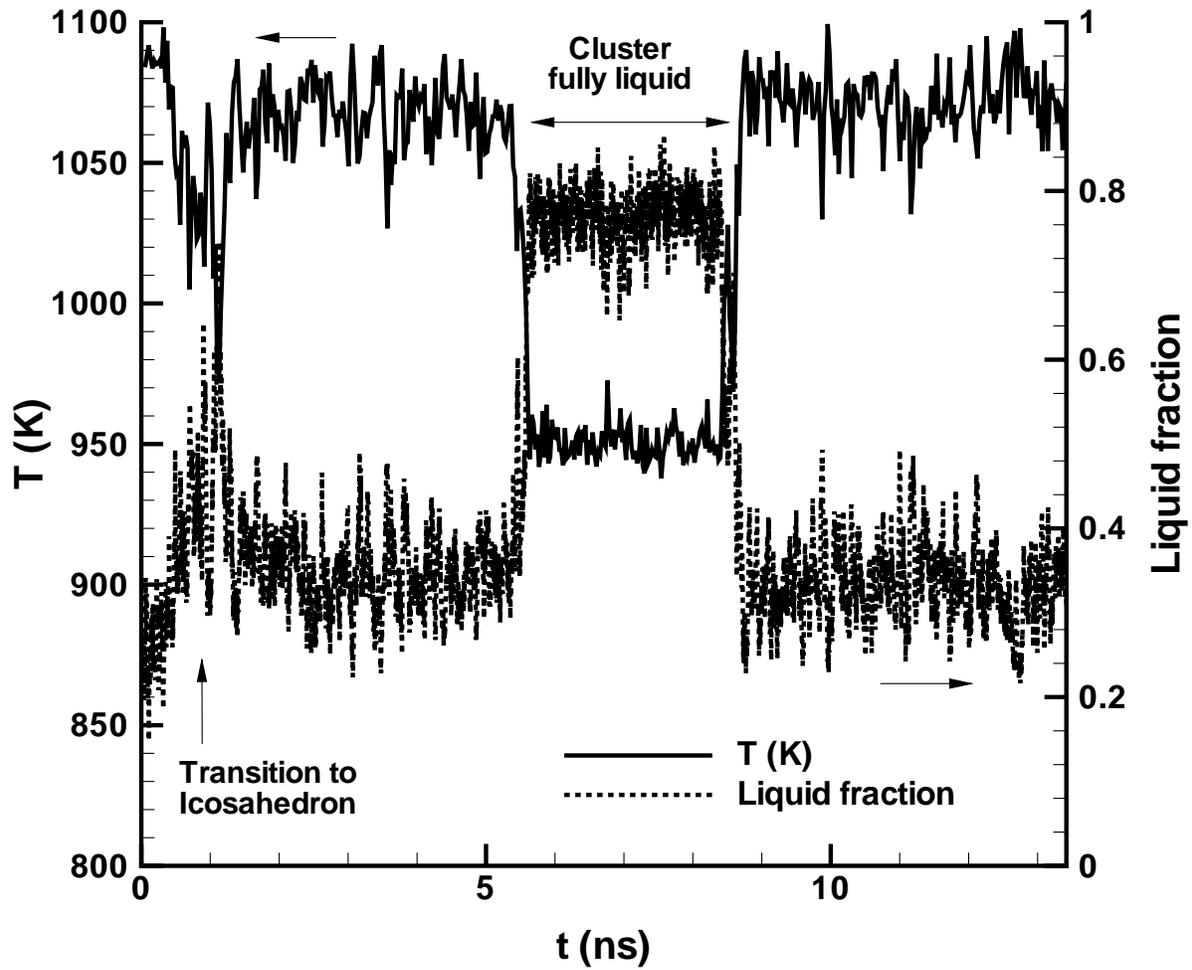}}
\caption{\label{fig5}The temperature and liquid fraction of an 586-atom cluster at $E=-3.278$ eV/atom initially prepared in a TO structure.}
\end{figure}

We also see a similar transition in the 586-atom TO. The caloric curve (not shown) reveals that solid-liquid coexistence begins at approximately $E=-3.290$ eV/atom, and melting occurs at $E=-3.270$ eV/atom. In figure~\ref{fig5} we show the time evolution of the temperature and the liquid fraction for a cluster prepared in the TO structure and then simulated at $E=-3.278$ eV/atom. At approximately 1 ns into the simulation, there is a large spike in the liquid fraction and temperature. This spike corresponds to the complete melting of the cluster, which subsequently recrystallizes into a partial icosahedral structure coexisting with the melt (much like that seen in the 887-atom cluster). Between 5.5 ns and 8.5 ns, the cluster again completely liquifies (although we calculate a liquid fraction of only 0.8, the distribution of mobilities has become unimodal indicating a fully liquid cluster \cite{Hendy05c}) and then again recrystallizes into the icosahedral structure. Thus, the cluster is exhibiting dynamic coexistence \cite{Honeycutt87} between a coexisting solid-liquid icosahedron and a fully liquid cluster.   

Similar behavior was observed in the other clusters studied here. In the 309-atom icosahedron, we saw dynamic coexistence between the partially melted icosahedron (the energetically stable structure at this size) and a fully liquid state. No unusual structural changes were observed in the solid. However, in the 1389-atom cluster we saw dynamic coexistence between the solid-liquid decahedral structure and the solid-liquid icosahedral structure, but did not see the 1389-atom cluster form a stable, long-lived icosahedral structure at any energy prior to melting. In the 2046-atom decahedron, we saw static coexistence prior to melting and a subsequent transition to a solid-liquid icosahedral structure as for the 887-atom decahedron. In the largest cluster examined here, the 2406-atom TO, we saw only static solid-liquid coexistence prior to melting, with no evidence for a transition to an icosahedral structure.   

All the transitions to icosahedral structures observed were preceded by a spike in the molten fraction of the cluster and corresponding drop in the cluster temperature. The excess melt that develops during these fluctuations then recrystallizes into partial icosahedral structures. In the particular case of the smaller 586-atom cluster, the solid appears to completely melt during this fluctuation and then recrystallizes as an icosahedron. In the larger decahedral clusters, only a fraction of the upper half of the decahedron remained solid during the fluctuation, and this then acts as a nucleus for recrystallization into the partial icosahedral structure. At the very least, these fluctuations in liquid fraction lower the energy barriers for the solid-solid transition. However, it is possible that these structural transitions are due to recrystallization kinetics rather than favorable entropy, since the recrystallization kinetics of the molten clusters undergoing a rapid quench typically favor the formation of icosahedral structures \cite{Baletto05}.      
In conclusion, we find that the fcc to icosahedral transition seen in MD simulations of a Pd cluster by Pundt et al \cite{Pundt02} was probably due to the instability of the cuboctahedral structure in the simulations. However, our simulations reveal solid-solid transitions in solid-liquid clusters prior to full melting that result from metastable fluctuations in the liquid fraction of the cluster. We note that this is a different mechanism to that the solid-solid transition seen in a partially melted Ni icosahedron \cite{Hendy05b} which was driven by facet-dependent wetting by the melt. These results demonstrate an important coupling between solid-liquid coexistence and solid-solid transitions in clusters near the melting point. Premelting features seen in cluster caloric curves may in fact be due to both processes.

\clearpage
\thispagestyle{empty}

\clearpage
\thispagestyle{empty}

\clearpage
\thispagestyle{empty}

\clearpage
\thispagestyle{empty}

\clearpage
\thispagestyle{empty}

\clearpage
\begin{figure}
\resizebox{\columnwidth}{!}{\includegraphics{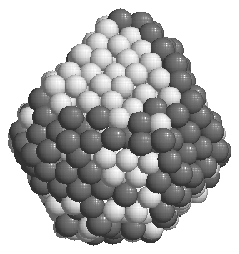} \includegraphics{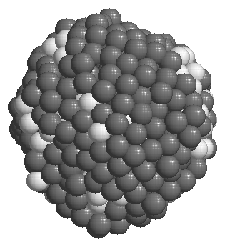} \includegraphics{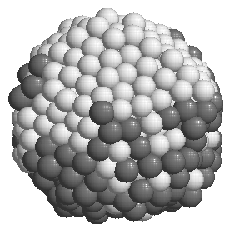}}
\caption{\label{fig6} The structure of the coexisting TO structure (left), the molten intermediate structure (center) and the coexisting icosahedral structure (right) in the 586-atom cluster. Atoms with a darker shade have been identified as liquid by their mobilities.}
\end{figure}
\end{document}